# A simple practical quantum bit commitment protocol

Muqian Wen

## Abstract

This paper devises a simple quantum bit commitment protocol that is just as easy to implement as any existing practical quantum bit commitment protocols but will be more secure. It will be infinitely close to being unconditionally fully secure although neither perfectly binding nor perfectly concealing.

## 1. Introduction

Bit commitment is an important topic in classical and quantum cryptography because it is related to many problems such as secure coin flipping and zero-knowledge proofs etc. There are classical bit commitment schemes [1]. But classical cryptography has a drawback in that they rely on limitations on the speeds of computers and algorithms to be secure which may not be future proof enough. This makes quantum security schemes interesting alternatives because they do not have to rely on computing power limitations. It has been shown that a certain kind of quantum bit commitment (QBC) protocols cannot be unconditionally secure [2,3]. Later, people found a kind of unconditionally secure quantum bit commitment protocol called relativistic quantum bit commitment and this has seen many papers in the top physics journal PRL [4-11]. But although relativistic quantum bit commitment protocols can be unconditionally secure, having to rely on special relativity would make them much less practical for actual uses. Thus, later some researchers would propose and demonstrate some practically secure non-relativistic quantum bit commitment protocols based on practical technological limitations [12-15]. The advent of practical quantum bit commitment protocols has rendered the issue of lack of unconditionally secure non-relativistic quantum bit commitment protocols less relevant because in real life it will always be the practical ones that matters, and it will never matter whether theoretical but practically unrealistic cheating methods exit or not. Practically secure quantum bit commitment schemes rely on technological limitations on quantum non-demolition measurements and long-term stable quantum memories to achieve security. It can be said that practically secure quantum bit commitment protocols will remain secure in the foreseeable future and thus it remains important to continue to study practically secure quantum bit commitment protocols. This paper would devise a very simple and practical quantum bit commitment protocol that is infinitely close to unconditionally secure. It will be more secure than existing practical practically secure protocols and yet is still as practical as possible for a quantum bit commitment protocol.

## 2. Theory

First, this paper will give a brief definition of the problem of bit commitment. Suppose Alice wants to tell Bob she has made a choice between two options, say 0 or 1, but she doesn't want Bob to know her choice until a later time. Meanwhile Bob needs to make sure Alice did have made a choice. So, they need to find a way out.

This paper will first reduce the problem of finding a fully secure bit commitment protocol into the problem of finding a partially binding and fully concealing protocol. Here binding means Alice cannot change her mind after committing and concealing means Bob cannot uncover the commitment before Alice reveals it. It can be shown that if there exists a partially binding and fully concealing protocol, then it will always be possible to construct an infinitely close to fully secure protocol simply by repeating the protocol many times. The proof is that while it will remain fully secure against Bob when the protocol is repeated multiple times, Alice will have to avoid getting caught on every repeat to cheat successfully, which will be effectively 0 when the number of repeats increases to infinity. Let us denote the cheating success rate of Alice in one repeat to be $p_A$ and the cheating success rate of Bob in one repeat to be $p_B$. Then the probability for Alice to successfully cheat $n$ times consecutively will be:

$$p_{An} = {p_A}^n$$

which will approach 0 as $n$ increases to infinity because $p_A$ is less than 1, and the chance for Bob to successfully cheat at least once is:

$$p_{Bn} = 1 - (1 - p_B)^n$$

which will remain 1 because $p_B$ equals 0. Actually $p_B$ does not need to be exactly 0 but only needs to be much closer to 0 than $1 - p_A$. This completes the proof. For practical purposes such infinitely close to fully secure protocol will be no different than being fully secure. Such protocol will also be efficient for quantum bit string commitments.

### 2.1. Single qubit conditionally secure protocol:

The simplest such practically secure protocol is a single qubit protocol which has been described in the literature before [12,13] which is the building block behind an existing practical quantum bit commitment protocol in this literature [12]. This is a conditionally partially binding and unconditionally fully concealing protocol. It is based on the first quantum key distribution protocol, the BB84 protocol. First, Bob sends to Alice a not long-term storable qubit in any one of the following states: $|0\rangle$, $|1\rangle$, $|+\rangle$, $|-\rangle$, which satisfy the following relations:

$$\langle 0|1\rangle = 0$$

$$\langle +|-\rangle = 0$$

$$|+\rangle = \frac{|0\rangle + |1\rangle}{\sqrt{2}}$$

$$|-\rangle = \frac{|0\rangle - |1\rangle}{\sqrt{2}}$$

If the commitment is 0, then Alice would use $01$ basis to measure it immediately after receiving the qubit. Likewise, if the commitment is 1, she would use $\pm$ basis. Later when Alice wanted to tell Bob her committed value, she would tell Bob her measurement result, which includes the measuring basis she used, as well as the measurement time.

It is obvious that this protocol is unconditionally fully concealing since Alice did not tell Bob anything. Although this is a non-relativistic quantum bit commitment protocol, the unconditional concealing property of this protocol can be conveniently proved by using special relativity. Special relativity dictates that information cannot be transmitted faster than light. But supposed that Bob will be able to infer the commitment values by Alice before the revealing stage through some kinds of measurements on his side if he had prepared the qubits in some kinds of entangled states, then this protocol can be used for instant transmission of information between two parties located very far away which is not allowed under special relativity. Specifically, Bob can first give a sufficiently large quantity of long term storable entangled qubits to Alice who is located far away and specify a time for Alice to write her messages by performing measurement on the entangled qubits on her side so that Bob can then retrieve these messages by performing measurement on the entangled qubits on his side. This way of exchanging information clearly violates special relativity because it happens essentially instantaneously even though the two parties are located arbitrarily far apart thus this is impossible. This completes the proof.

This protocol is only conditionally binding because Alice will have a 100% cheating success rate by simply deferring the measurement of the qubit until last minute if there are no technological limitations. But if there is a technological limitation that the qubit is not long-term storable and has to be measured immediately upon receipt, then Alice will only be able to achieve a 75% cheating success rate in such a condition. Thus, this is a conditional practically partially binding protocol. The 75% cheating success rate for Alice is very trivial to see. First, when Alice wanted to change her mind, she would have to guess a measurement outcome with a different measurement basis. There will be a 50 percent chance that the new measuring basis is incorrect. Then whatever outcome Alice guesses will not matter because all guesses are equally plausible. However, if the new basis happens to be correct, then there will be a 50 percent chance that the guessed result by Alice is impossible. In this scenario Bob will be able to know that Alice has cheated on him. Thus overall, Alice will have a 25 percent chance of being caught cheating. This completes the proof. To put it mathematically, suppose that the probability that the measurement basis the cheating Alice pretend to choose is the same as the basis that the qubit is actually prepared in by Bob is $p_r$ and the probability of the guessed measurement outcome that Alice pretended to obtain contradicting with the actual state that the qubit were prepared in is $m_w$. Then the probability of Alice being caught in cheating will be:

$$p_r m_w = 0.5 * 0.5 = 0.25$$

### 2.2. Multi qubit conditionally secure protocol:

Next comes the innovation of this paper, instead of simply repeating this protocol many times to create a new conditionally fully secure protocol like the previous literature did [12,13], this paper would create a novel $n$ qubit protocol by chaining multiple ones of this single qubit protocol. Instead of Alice keeping the measurement result to herself, Bob sends Alice a second qubit to commit the measurement result of the first qubit and likewise for all subsequent qubits until the n-th qubit, at which time she would keep the measurement result to herself. To be more specific, if the measurement result by Alice is either $|0\rangle$ or $|+\rangle$, she must use $01$ basis for the next qubit and likewise for the other case. The two parties can use classical communication channels to make sure that each qubit is received successfully. The revealing stage would be the same as previous protocols which is that Alice tells Bob all the measurement results.

Next, this paper will derive the measurement basis of this protocol which is not as apparent as the previous protocols. The $n$ number of qubits in this protocol can be considered as a single quantum system with $2^{n+1}$ number of possible measurement outcomes. Every possible measurement outcome is an eigenstate of one of the measuring bases in this protocol. The possible measurement outcome or eigenstates can be expressed as:

$$|\psi_e\rangle = \prod_{i=1}^{n} |b_i, q_i\rangle_i \qquad i = 1,2, \dots n$$

where $|\psi_e\rangle$ is the eigenstate of this $n$ qubit quantum system, $b_i$ represents the basis of the $i$-th qubit which is either $01$ or $\pm$ basis while $q_i$ determines if the state of the qubit is one of the $|0\rangle$, $|+\rangle$ or one of the $|1\rangle$, $|-\rangle$. If we consider $|0\rangle$, $|+\rangle$ as positive states and $|1\rangle$, $|-\rangle$ as negative states, then $q_i$ determines whether the qubit is in positive or negative state. Likewise, we can also regard the $01$ basis as positive basis and the $\pm$ basis as negative basis out of convenience. The value of $b_i$ is determined by the state of the previous qubit $q_{i-1}$ by essentially a sign function:

$$b_i = sgn(q_{i-1})$$

If $q_{i-1}$ is a positive state, then $b_i$ will be $01$ basis and likewise for the other case. It can be proved that there are 2 measuring bases in this protocol with the first basis being when the first qubit is measured in 01 basis and the second basis being when the first qubit is measured in $\pm$ basis. To prove this, it would be sufficient to prove that the eigenstates of each basis are orthogonal to each other and that all non-eigen states of this $n$ qubit system can be rewritten as the linear combinations of these eigenstates. The latter is trivial to see. For the first one, it is also trivial to see that the following is true:

$$\langle \psi_e' | \psi_e \rangle |_{b_0'=b_0} = 0$$

Thus, this completes the proof. Having figured out the measurement basis in this n qubit protocol, we can then discuss the properties of this protocol.

This novel n qubit protocol has the same security property as the original single qubit version which is conditionally partially binding and unconditionally fully concealing. The unconditionally fully concealing property can be proved using the same argument in the previous section for the single qubit protocol. The conditionally partially binding property can be proved in a similar way as well. If there are no technological limitation Alice will have a 100% cheating success rate by simply delaying the measurement. If there is a technological limitation that the qubits are not long-term storable and therefore must be measured immediately, then Alice will be able to achieve a maximum 75% cheating success rate. The proof is the following. First, the best way for Alice to cheat is to lie in way that only the first qubit measurement result is lied so that Bob will only be able to use the first qubit to help determine if Alice is cheating. In this case, the condition for Alice to be caught in cheating is that both the first qubit and the second qubit were measured in wrong basis so that when Alice tried to lie about the first qubit measurement result, the lied result will be an impossible outcome for the qubit. The probability for both the first qubit and the second qubit to be measured in wrong basis is:

$$p_{1bw} p_{2bw} = 0.5 * 0.5 = 0.25$$

Where $p_{1bw},\ p_{2bw}$ is the probability of the basis of each qubit being wrong respectively. This completes the proof.

Although this protocol is no more secure than previous existing protocols, it will have an advantageous property in that most of the qubits will be measured in correct bases. This probability is governed by the negative binomial distribution which dictates that if Alice chose a wrong basis on the first qubit, then on average by the third qubit all subsequent qubits will be measured with correct basis. By comparison only half qubits will be measured in correct basis in existing protocols. Thus, this protocol is more efficient. One scenario where this property can be useful is if Alice tries to delay measurement. Then because of the delay, many of the qubits would have decayed and as a result the measurement outcome can be wrong even if the measuring basis is correct which would reveal to Bob that the measurement was delayed. Since in this protocol almost all qubits will be measured in correct basis as opposed to only half in other protocols, it will be more efficient.

### 2.3. Unconditionally secure protocol:

Finally, this paper further evolves this n qubit protocol by requiring that Alice tells Bob the measurement outcome of the last qubit immediately after measurement. This protocol will be unconditionally partially binding and unconditionally infinitely close to fully concealing. It is unconditionally partially binding because Alice will no longer be able to cheat by delaying measurement. If she tries to delay measurement, then the success rate will be only 25% because she will have to guess the measurement result of the last qubit. Note that it is not necessary to require Alice to reveal the full measurement result of the last qubit to prevent delay measurement. If instead Alice is only required to tell Bob whether the measured result of the last qubit is a positive or negative state, then the cheating success rate by delaying will still be only 50%. Without delay measurement the cheating success rate for Alice will be 75%. This can be proved using the same arguments as the previous protocol in the previous section. This protocol cannot have a better cheating success rate for Alice than the previous protocol because Alice needs to reveal more information to Bob in this protocol than the previous one.

This protocol will be unconditionally infinitely close to fully concealing. It is no longer perfectly concealing like the previous protocol because now Alice is required to reveal some information to Bob. Assuming that the qubits are prepared in an eigenstate, the condition for Bob to find out the commitment value by Alice with certainty is that the last qubit measurement outcome is incorrect which can only happen if the first qubit happens to be measured with the wrong basis and all subsequent measurement results also align to make the chosen measuring basis of all qubits wrong and the probability for this to happen is:

$$p_B = \prod_{i=1}^{n} p_{ibw} = \left(\frac{1}{2}\right)^n$$

Which will approach 0 as $n$ reaches infinity. Thus, this protocol will be infinitely close to fully concealing. The last remaining question is whether the cheating success rate can be improved by not preparing the $n$ qubits in eigen states. The proof of this concealing property is the following. The $n$ qubits can be prepared in a pure state or an entangled state with or without some other qubits, although in practice entangled states will likely be impossible to make due to technological limitations. Any quantum state can be written as a linear combination of eigen states. There are two bases so a state can be written in two ways. Consider the most general case where the $n$ qubits can be entangled with some additional qubits which Bob will do some measurements on. Then a state can be expressed as:

$$|\psi\rangle = \sum_{i=1}^{2^n} \sum_{j=1}^{m} C_{ij} |\varphi_i\rangle |\phi_j\rangle = \sum_{i=1}^{2^n} \sum_{j=1}^{m} C'_{ij} |\varphi'_i\rangle |\phi'_j\rangle$$

Where $\psi$ represents the total quantum state of the $n$ qubits and the additional qubits, $\varphi_i$ is the $i$-th eigen state of the $n$ qubits in the first measuring basis, $m$ is the number of eigen states of the additional qubits, $\phi_j$ is the $j$-th eigen state of the additional qubits, $C_{ij}$ is a constant, $\varphi'_i$ is the eigen state in the second measuring basis. The measurement operation on the additional qubits can depend on the prepared state of the $n$ qubits and the measurement outcome of the last qubit so $\phi_j$ can also depend on them. An eigen state in one basis can be expanded in another basis as:

$$|\varphi'_i\rangle |\phi'_j\rangle = \sum_{k=1}^{n-1} \frac{(-1)^{g_k}}{2^k} |\varphi_{ik}\rangle |\phi'_j\rangle + \frac{(-1)^{g_{n-1}}}{2^n} |\varphi_{i+}\rangle \sum_{h=1}^{m} C_h^+ |\phi_h^+\rangle + \frac{(-1)^{g_n}}{2^n} |\varphi_{i-}\rangle \sum_{t=1}^{m} C_t^- |\phi_t^-\rangle$$

Where $\varphi_{ik}$ is an eigen state in the first basis that has same qubit states as $\varphi'_i$ for the last $n-k$ number of qubits, $\varphi_{i+}$ and $\varphi_{i-}$ are two eigen states that have different individual qubit state than $\varphi'_i$ for all $n$ qubits with $\varphi_{i+}$ being the one where the last qubit is of a positive state and likewise for $\varphi_{i-}$, $g_k$ denotes the number of minus signs encountered during expansion, $\phi_h^+$ and $\phi_t^-$ are corresponding eigen states of the additional qubits which may or may not be the same as $\phi'_j$ depending on how Bob chose to measure them based on the last qubit measurement outcome by Alice. $C_h^+$ and $C_t^-$ are the corresponding constants. In the eigen states of the first term of this equation, the last qubit state and the state of additional qubits are the same as the original eigen state on the left side of the equation. The second and third term eigen states can only result from Alice choosing their corresponding measurement bases and collapsing into these states will reveal to Bob which commitment value Alice has chosen. But the probability to collapse to these states will infinitely decrease to 0 as $n$ increases to infinity. If Alice is only required to reveal the sign of measured state of the last qubit, then one of the second or third term can be placed into the first term and further reduce the cheating success rate of Bob by another factor of 2. Since any non-eigen state is just a linear combination of eigen states, preparing the $n$ qubits in a non-eigen state will not improve the chance that these qubits collapse into the second and third term eigen states after measurement. And it is also apparent from the equation that entangling the $n$ qubits with some additional qubits will not help Bob improve cheating success rate either. Thus, this protocol is infinitely close to fully unconditionally concealing. This completes the proof. Simply repeating this protocol many times can construct an unconditionally infinitely close to fully secure protocol. And when committing a string of bits each repeat can also correspond to different bits to significantly improve efficiency.

## 3. Discussion

This paper has successfully devised a practical infinitely close to unconditionally fully secure quantum bit commitment protocol. The significance of this protocol is that it brings practicality and unconditional security together for the first time. Compared to existing practical protocols it is much closer to unconditional security while compared to existing unconditionally secure relativistic protocols it is much more practical because it does not rely on special relativity. It is just as practical to implement as any existing practical quantum bit commitment protocols in the literature while reaching security levels of unconditionally secure protocols.

The principle of practically secure quantum bit commitment protocols is that cheating would require much more difficult and unrealistic technologies than simply not cheating and thus can be effectively secure. Photon can be considered such a kind of qubit carrier that non-demolition measurement and long-term storage are beyond technological reach in the foreseeable future and practical quantum bit commitment schemes implemented using photons has been demonstrated. Thus, practical quantum bit commitment protocols can remain relevant for the foreseeable future. Practical quantum bit commitment protocols can be implemented cheaply with relatively simple technologies while requiring unrealistic technologies to break. The original unsecure quantum bit commitment protocol is actually practically secure with current technologies but will also be impossible to implement with current technologies. There are probably few application cases that would need unconditional security at all costs, thus practical protocols are probably more useful in practical sense than relativistic protocols. Real world applications will most likely also need to commit more than one bit which unpractical protocols never considered.

## 4. Conclusion

This paper has devised a simple and very practical infinitely close to unconditionally fully secure quantum bit commitment protocol. The significance of it is that it brings practicality and unconditional security together for the first time. It is hoped that this study can be useful for developing better quantum bit commitment protocols and for other quantum cryptography research.

### Data statement

No new data is generated in this paper.

### Conflicts of interest

There are no known conflicts of interests.